\documentclass[a4paper,11pt]{article}
\usepackage{pos}
\usepackage{graphicx}
\usepackage[caption=false]{subfig}
\usepackage{float}
\usepackage{wrapfig}
\usepackage{multirow}
\usepackage{longtable}
\usepackage{hyperref}
\title{FAIR Principles for data and AI models in high energy physics research and education }

\author*[1]{Avik Roy}

\affiliation{National Center for Supercomputing Applications, University of Illinois at Urbana-Champaign \\
1205 Clark St, Urbana, Illinois 61801, United States of America}
\note{on behalf of the FAIR4HEP collaboration: \href{https://fair4hep.github.io}{https://fair4hep.github.io}}
\emailAdd{avroy@illinois.edu}

\abstract{
In recent years, digital object management practices to support findability, accessibility, interoperability, and reusability (FAIR) have begun to be adopted across a number of data-intensive scientific disciplines. These digital objects include datasets, AI models, software, notebooks, workflows, documentation, etc. With the collective dataset at the Large Hadron Collider scheduled to reach the zettabyte scale by the end of 2032, the experimental particle physics community is looking at unprecedented data management challenges. It is expected that these grand challenges may be addressed by creating end-to-end AI frameworks that combine FAIR and AI-ready datasets, advances in AI, modern computing environments, and scientific data infrastructure. In this work, the FAIR4HEP collaboration explores the interpretation of FAIR principles in the context of data and AI models for experimental high energy physics research. We investigate metrics to quantify the FAIRness of experimental datasets and AI models, and provide open source notebooks to guide new users on the use of FAIR principles in practice.
}

\FullConference{%
  41st International Conference on High Energy physics - ICHEP2022\\
  6-13 July, 2022\\
  Bologna, Italy
}

\begin{document}
\maketitle

\section{Introduction}
\vspace{-5pt}
In order to make datasets findabile, accessibile, interoperable, and reusable (FAIR), a set of data principles have been defined so that scientific datasets could be readily reused by both humans and machines~\cite{wilkinson2016fair}. Originally envisioned for preservation of scientific datasets, the FAIR principles have been interpreted in the context of different kinds of digital objects, including research software~\cite{fair4rs}, notebooks~\cite{richardson2021user}, custom digital libraries~\cite{neubauer2022making}, and machine learning (ML) models~\cite{katz2021working, ravi2022fair}. This work summarizes the multifaceted interpretation and applications of the FAIR principles explored by the FAIR4HEP collaboration in the context of High Energy Physics (HEP).

\section{FAIR principles for HEP data}
\begin{wrapfigure}{R}{0.45\textwidth}
  \begin{center}
    \centering
    \includegraphics[width=0.42\textwidth]{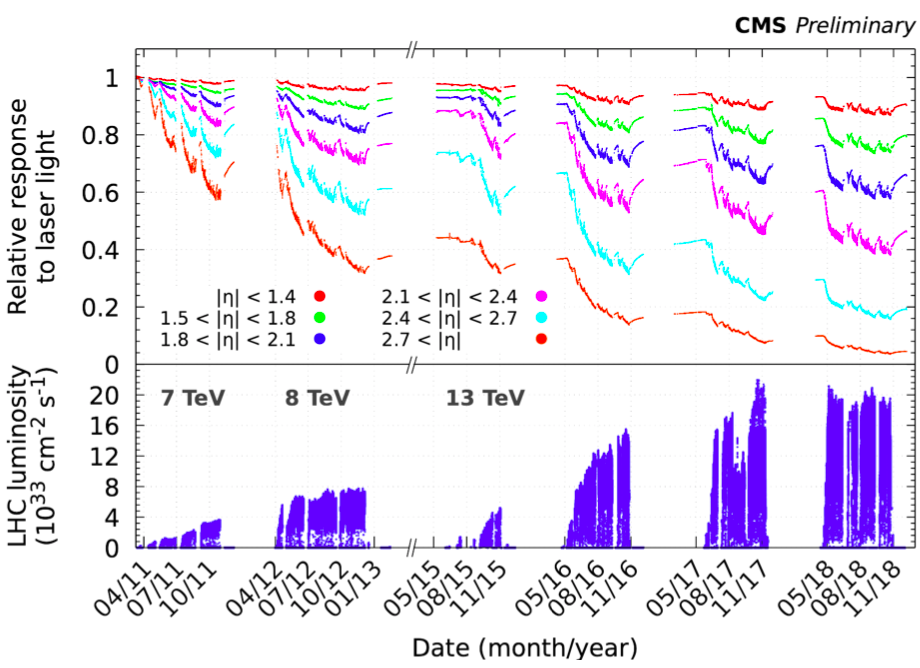}
  \end{center}%
\caption{Transparency of CMS ECal crystals in different regions of pseudorapidity $(\eta)$  during Run-2. 
This image is taken from Ref.~\cite{CMS-DP-2019-005}
}
\label{fig:CMS-Ecal}
\end{wrapfigure}

\vspace{-5pt}
Interpretation and application of FAIR principles for HEP datasets has been explored in Ref.~\cite{chen2022fair}. Taking the publicly available $H \to b\bar{b}$ dataset from CMS Open Data portal,
this work performs a thorough evaluation of the FAIR-readiness of this dataset using domain-agnostic metrics. Having the established the FAIRness of this dataset, this work demonstrates how adhering to these principles allow this dataset to be AI-ready. This dataset has been made available in formats widely used by the broader ML community. Pedagogical examples of usage of this dataset shows implementation of cutting edge methods in ML for jet classification. Guided by the FAIR principles, the FAIR4HEP collaboration has published multiple datasets for a variety of problems in HEP.
\vspace{-6pt}
\begin{itemize}
    \item \textbf{Super Cryogeinc Dark Matter Search (Super CDMS) dataset}~\cite{cdms-data} was obtained from a prototype detector (Figure~\ref{fig:CDMS-det}) for recording phonon scattering from an interacting particle. With six operating channels (Figure~\ref{fig:det-3d}), this detector operates at 30~mK and records phonon pulse amplitude as a function of time in each channel (Figure~\ref{fig:waveform}). This dataset consists of timing information about these pulses for 7000 individual measurements taken over 13 different impact locations on the detector. This dataset is being used to reconstruct the impact location based on these timing information and to obtain generative models for simulation of particle intraction.
    
\begin{figure}[h]
\centering
\subfloat[]{
\includegraphics[width=0.29\textwidth]{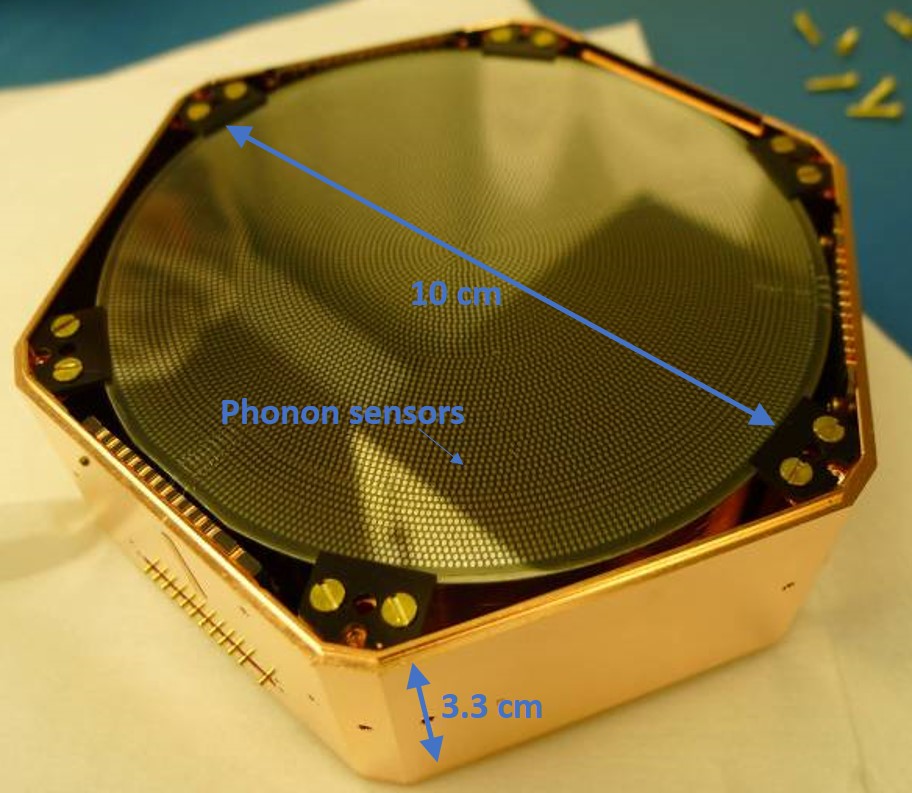}
\label{fig:CDMS-det}            
}
\subfloat[]{
\includegraphics[width=0.3\textwidth]{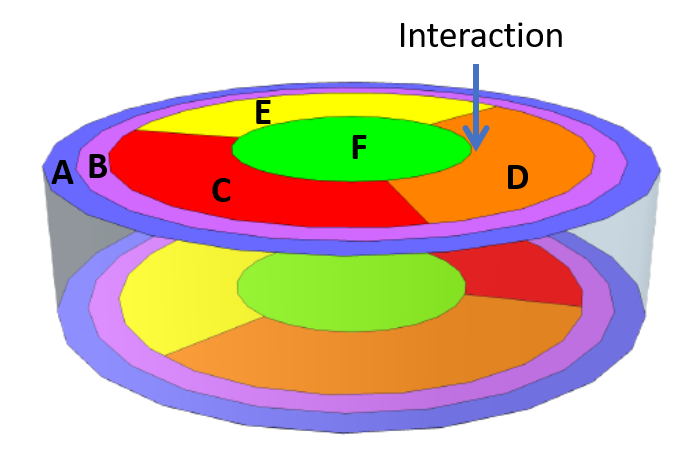}
\label{fig:det-3d}            
}
\subfloat[]{
\includegraphics[width=0.3\textwidth]{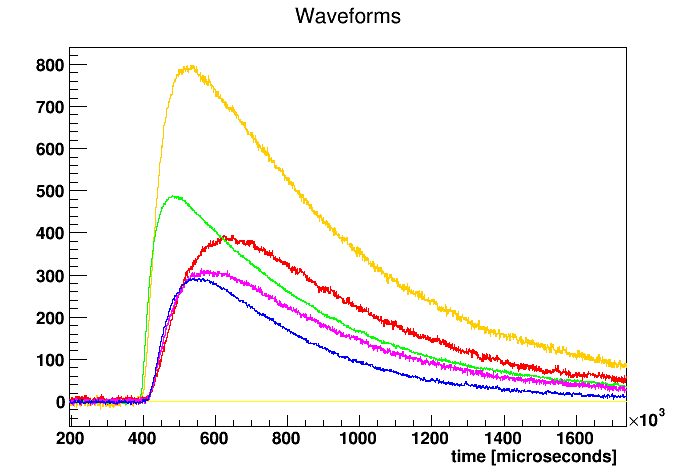}
\label{fig:waveform}            
}
\caption{\protect\subref{fig:CDMS-det} A Super CDMS detector prototype and \protect\subref{fig:det-3d} illustration of its different channels. \protect\subref{fig:waveform} shows the waveforms recorded by different channels for an example particle interaction. These images are taken from https://fair-umn.github.io/FAIR-UMN-CDMS/
}
\label{fig:CDMS}
\end{figure}
\vspace{-6pt}
    \item \textbf{Laser Response from Electromagnetic Calorimeter (ECal) crystals at CMS}~\cite{cms-ecal} is obtained from the thousands of calibrations performed during the course of Run-2 between 2016 and 2018. The ECal calorimeter at CMS consists of almost 758k lead tungstate ($PbWO_4$) crystals. Exposure to prolonged radiation during data taking modifies the transparency of these crystals, which is partially recovered when such exposure is taken away (Figure~\ref{fig:CMS-Ecal}). This dataset contains information about the Ecal crystal's laser response and can be used to model this behavior as a function of integrated luminosity to predict future behavior.

\end{itemize}

\section{FAIR Principles for AI Models}
\begin{figure}[!h]
  \begin{minipage}[c]{0.65\columnwidth}
    \centering
    \includegraphics[width=\columnwidth]{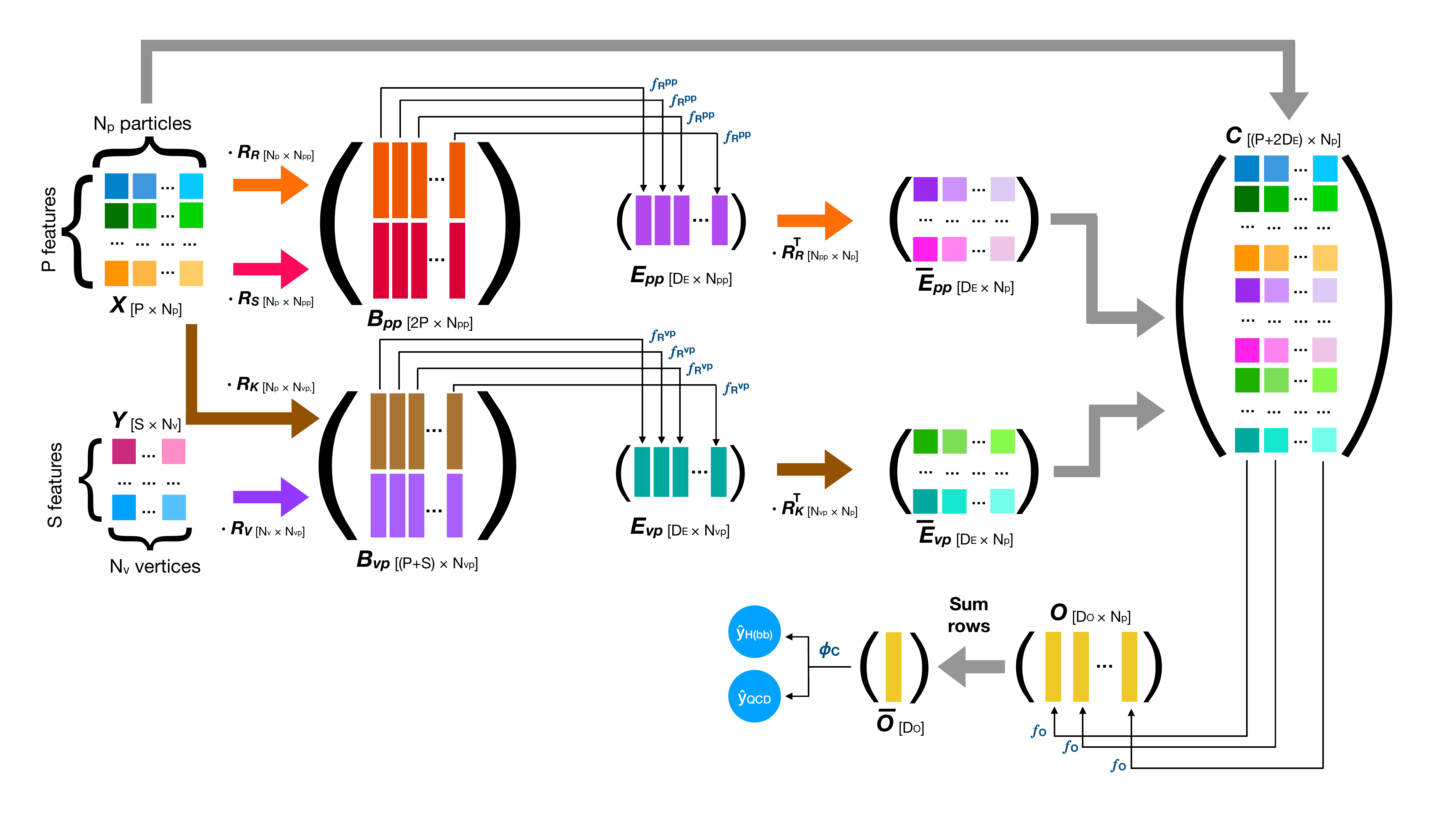}
  \end{minipage}%
  \begin{minipage}[c]{0.150\columnwidth}
    \centering
\begin{tabular}[b]{ |c|c|c| }
  \hline
  \multicolumn{2}{|c|}{Default IN hyperparameters} \\
  \hline \hline
  $(P, N_p, S, N_v)$ &  $(30, 60, 14, 5)$ \\ \hline
  Hidden layers &  3\\ \hline
 Hidden layer   &  \multirow{2}{*}{60}\\ 
 dimension & \\ \hline
  $(D_e, D_o)$ &  (20, 24)\\ \hline
  Activation &  ReLU\\ \hline%
\end{tabular}
\end{minipage}
\caption{A Schematic diagram of the network architecture and dataflow in the IN model. This image is taken from Ref.~\cite{IN}. The choice of model hyperparameters and input data dimensions for the baseline model is given in the accompanying table.}
\label{fig:IN-arch}
\end{figure}

\vspace{-5pt}
Application of the FAIR principles for AI models requires incorporating their dependence on computing environments into the interpretation of FAIR principles. We use the Interaction Network (IN) model~\cite{IN} for classification of jets from the aforementioned $H\to b\bar{b}$ dataset.
Figure~\ref{fig:IN-arch} shows the IN model architecture and tabulates the default hyperparameters and data dimensions. This network is built to train on graph data structure whose nodes comprise of $N_p$ particle tracks, each with $P$ features, and $N_v$ secondary vertices, each with $S$ features, associated with the jet. The physical description of each feature is given in Appendix C of ref.~\cite{IN}. It creates a fully connected directed graph  with $N_{pp} = N_p(N_p - 1)$ edges for the particle tracks. A separate graph with $N_{vp} = N_vN_p$ generates all possible connections between the particle tracks and the secondary vertices. 

The node level features for the track-track (track-vertex) graph are
transformed to edge level features via a couple of interaction matrices, 
transformed via fully connected NNs, 
%
and finally transferred back to the track-level representations by the interaction matrices. 
The trainable dense MLP creates the post-interaction internal representation that are summed over the tracks and then linearly combined to produce a two-dimensional, softmax-transformed output to predict individual class probabilities.  

\subsection{Interoperability of the IN mode}

To inspect that the model can infer under multiple deep learning frameworks, we convert the IN model to \texttt{ONNX}~\cite{onnx} format and then to \texttt{TensorRT}~\cite{trt} engine from the \texttt{ONNX} format model. 
The \texttt{ONNX} format is designed to help the model become portable between the commonly used deep learning formats like \texttt{PyTorch, TensorRT, TensorFlow} etc. 
We also quantitatively evaluate the inference performance before and after the conversion by measuring inference accuracy, running time, and Area under the Curve (AUC) for the Receiver Operating Characteristic (ROC) curve.

We separately infer three test groups using the \texttt{PyTorch} and \texttt{ONNX} model with a batch size of  128. With all three methods of inference, the model's performance in terms of accurancy and the ROC-AUC score was identical. When comparing the average time needed for inference per batch, the \texttt{PyTorch} model took 1.226~ms, while inference with \texttt{ONNX} and \texttt{TensorRT} models took 12.64 ans 12.31~ms respectively. This order of magnitude difference can be understood from the data transfer between the GPU device hosting these models and the actual host of the data. 
The running time/epoch refers to the time used to run one batch, including the data transfer between the two sides (device, host) and the inference part in the GPU device. 
When we increase the batch size from 1 to 128, the running time becomes larger, because the time used for data transfer also increases. The running time difference between the \texttt{ONNX} and \texttt{TensorRT} models can be attributed to the differences in hardware acceleration. 

\begin{wrapfigure}{R}{0.45\textwidth}
  \begin{center}
    \centering
    \includegraphics[width=0.42\textwidth]{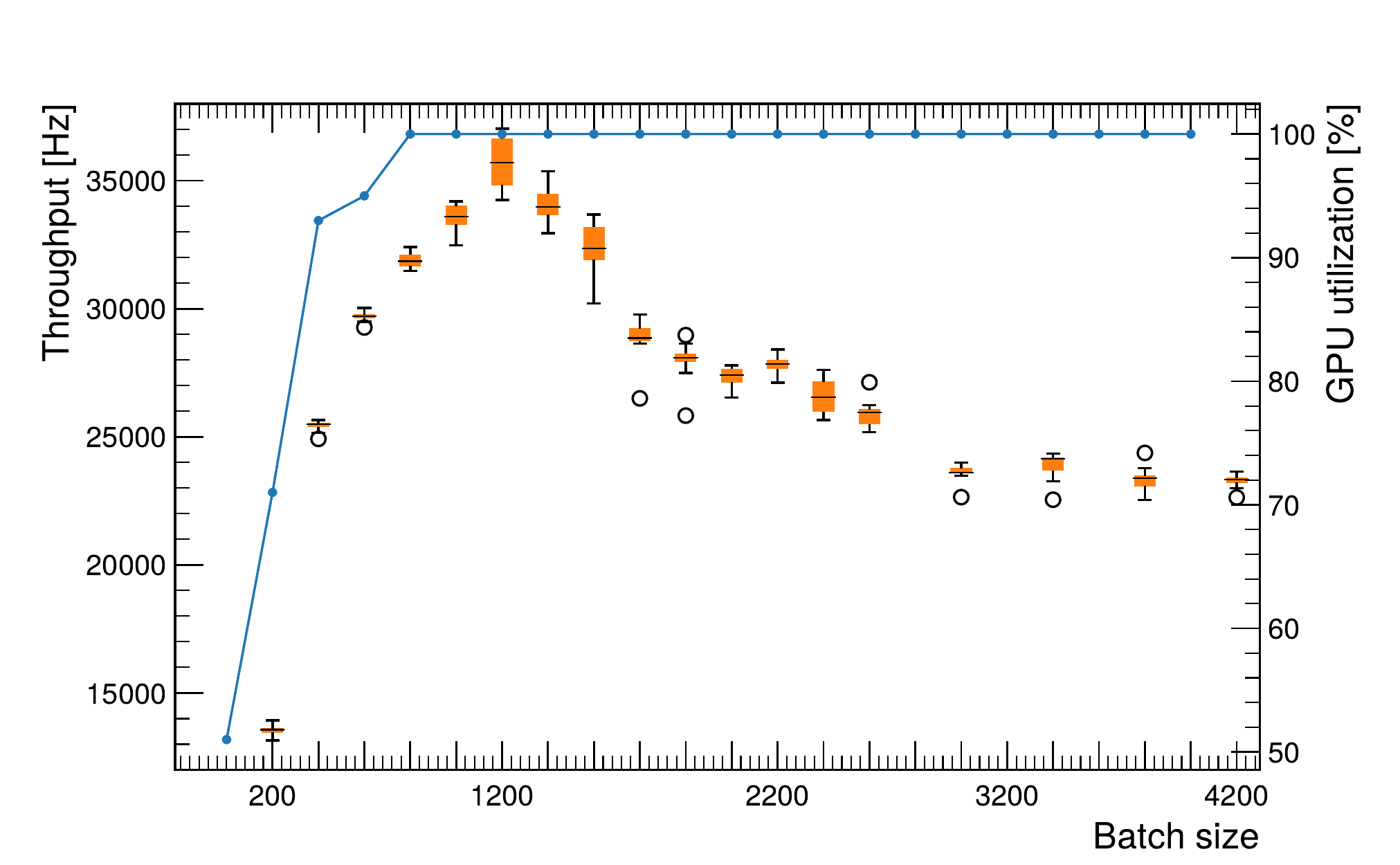}
  \end{center}%
\caption{GPU utility and throughput for inference with \texttt{TensorRT} engine with different batch size 
}
\label{fig:util-throughput} 
\end{wrapfigure}

Here we also test whether different batch size will influence the inference results using the model. 
We record the performance of the GPU utility and throughput, i.e the number of inferences performed per second, with the change of batch size in \texttt{TensorRT} format. The model's performance was stable across batch sizes.
Figure~\ref{fig:util-throughput} shows the results. GPU utility steadily increases up to a batch size of 1000 and saturates at 100\% beyind that. 
The performance of throughput also matches with the GPU utility, it reaches maximum when batch size reaches around 1200, which is the almost same period when the GPU utility is fully occupied.



\subsection{Streamlining FAIR AI Development}
As AI models are often developed as an intricate software repository, an automation tool to allow a streamlined organization of such repositories has been developed. 
Cookiecutter Data Science~\cite{cookiecutter_data_science} provides one such template, specifically oriented at data science projects.
It consists of a logical, reasonably standardized, but flexible project structure hosted on GitHub for performing and sharing data science work.
We took inspiration from this and created a fork of this template generator, called \texttt{cookiecutter4fair}~\cite{cookiecutter4fair}, with additional features to promote the adoption of our FAIR principles.
It allows structural organization of machine learning model repositories, development of containers, and publishing the codebase with persistent identifiers like digital object identifier (DOI). It includes comprehensive instructions to build the necessary environments and allows coherently incorporating notebooks for various studies. This tool has been used to restructure the codebase for the IN model~\cite{hbb_interaction_network} and include notebooks that facilitate studies on its interoperbility and explainability.
\vspace{-5pt}
\section{Pedagogical aspects of FAIR in HEP}
\vspace{-5pt}
FAIR principles can facilitate education in data science and machine learning. Developing classroom contents aligned with the FAIR principles can amplify their reach and effectiveness. On the other hand, introducing students with FAIR principles and their interpretation for data and ML can help standardize modern practices in digital object management.
\vspace{-6pt}
\begin{itemize}
    \item \textbf{Pedagogical Introduction to FAIR}\footnote{\href{https://github.com/yorkiva/FAIR-Exercises}{https://github.com/yorkiva/FAIR-Exercises}} is a series of notebooks developed based on the previously mentioned Super CDMS dataset. It contains an introduction to FAIR principles, evaluation of FAIRness of datasets, and builds up ML models, both traditional and neural network based, and demonstrates their organization and management in line with the FAIR principles. These notebooks include the analysis of the impact location estimation problem with the CDMS dataset using regularized linear regression, principal component analysis, deep neural netowrks, and variational autoencoders.
    \vspace{-6pt}
    \item \textbf{Data Science for Physics}\footnote{\href{https://github.com/MIT-8s50/course}{https://github.com/MIT-8s50/course}} is a course designed to introduce modern concepts of machine learning via data analysis problems in HEP. This course was originally developed as an online substitute to the junior lab at MIT during the COVID pandemic and will be offered as a full, independent class starting from Spring 2023. It includes as series of lectures on different topics of physics, statistics, and data analysis and requires analysis of physics data from LIGO, CMS, and CHIME experiments.  An open source version of this course will also be featured on the MITx\footnote{\href{https://mitxonline.mit.edu}{https://mitxonline.mit.edu}} platform. 
\end{itemize}

\section{Conclusion}
\vspace{-5pt}
 With a view to inspiring the modern community-wide standards for preservation and management of digital objects, FAIR4HEP is developing HEP-specific interpretation of FAIR and active implementation by developing FAIR datasets, models, and tools. This work summarizes the ongoing efforts of the project along these ventures.
\vspace{-5pt}
\section*{Acknowledgements}
\vspace{-5pt}
This work was supported by the FAIR Data program of the U.S. Department of Energy, Office of Science, Advanced Scientific Computing Research, under contract number DE-SC0021258.




\bibliography{PROC-ICHEP}

\providecommand{\href}[2]{#2}\begingroup\raggedright\begin{thebibliography}{10}

\bibitem{wilkinson2016fair}
M.D.~Wilkinson, M.~Dumontier, I.J.~Aalbersberg, G.~Appleton, M.~Axton, A.~Baak
  et~al., \emph{The {FAIR} guiding principles for scientific data management
  and stewardship},
  \href{https://doi.org/https://doi.org/10.1038/sdata.2016.18}{\emph{Scientific
  data} {\bfseries 3} (2016) 1}.

\bibitem{fair4rs}
M.~Barker, N.P.~Chue~Hong, D.S.~Katz, A.-L.~Lamprecht, C.~Martinez-Ortiz,
  F.~Psomopoulos et~al., \emph{Introducing the fair principles for research
  software},
  \href{https://doi.org/https://doi.org/10.1038/s41597-022-01710-x}{\emph{Scientific
  Data} {\bfseries 9} (2022) 1}.

\bibitem{richardson2021user}
R.A.~Richardson, R.~Celebi, S.~Van Der~Burg, D.~Smits, L.~Ridder, M.~Dumontier
  et~al., \emph{User-friendly composition of {FAIR} workflows in a notebook
  environment},
  \href{https://doi.org/https://doi.org/10.1145/3460210.3493546}{\emph{Proceedings
  of the 11th on Knowledge Capture Conference} (2021) 1}.

\bibitem{neubauer2022making}
M.S.~Neubauer, A.~Roy and Z.~Wang, \emph{Making digital objects fair in high
  energy physics: An implementation for universal feynrules output (ufo)
  models},
  \href{https://doi.org/https://doi.org/10.48550/arXiv.2209.09752}{\emph{arXiv
  preprint arXiv:2209.09752} (2022) }.

\bibitem{katz2021working}
D.S.~Katz, F.E.~Psomopoulos and L.J.~Castro, \emph{Working towards
  understanding the role of {FAIR} for machine learning},
  \href{https://doi.org/10.4126/FRL01-006429415}{\emph{Proceedings of the 2nd
  Workshop on Data and research objects management for Linked Open Science}
  (2021) 1}.

\bibitem{ravi2022fair}
N.~Ravi, P.~Chaturvedi, E.~Huerta, Z.~Liu, R.~Chard, A.~Scourtas et~al.,
  \emph{Fair principles for ai models with a practical application for
  accelerated high energy diffraction microscopy},
  \href{https://doi.org/https://doi.org/10.1038/s41597-022-01712-9}{\emph{Scientific
  Data} {\bfseries 9} (2022) 1}.

\bibitem{chen2022fair}
Y.~Chen, E.~Huerta, J.~Duarte, P.~Harris, D.S.~Katz, M.S.~Neubauer et~al.,
  \emph{A {FAIR} and {AI}-ready {Higgs} boson decay dataset},
  \href{https://doi.org/https://doi.org/10.1038/s41597-021-01109-0}{\emph{Scientific
  Data} {\bfseries 9} (2022) 1}.

\bibitem{cdms-data}
M.~Fritts and T.~Li, \emph{{CDMS}-dataset},  2021.
\newblock 10.34740/kaggle/dsv/2660709.

\bibitem{cms-ecal}
B.~Joshi and R.~Rusack, \emph{Laser response in {ECAL} crystals in {CMS}
  detector},  Mar, 2022.
\newblock 10.5281/zenodo.6394777.

\bibitem{CMS-DP-2019-005}
{\scshape CMS} collaboration, \emph{{CMS ECAL Response to Laser Light
  (CERN-CMS-DP-2019-005)}},  2019.

\bibitem{IN}
E.A.~Moreno, T.Q.~Nguyen, J.-R.~Vlimant, O.~Cerri, H.B.~Newman, A.~Periwal
  et~al., \emph{Interaction networks for the identification of boosted h→ b b
  decays},
  \href{https://doi.org/https://doi.org/10.1103/PhysRevD.102.012010}{\emph{Physical
  Review D} {\bfseries 102} (2020) 012010}.

\bibitem{onnx}
J.~Bai, F.~Lu, K.~Zhang et~al., ``{ONNX: Open Neural Network Exchange}.''
  \url{https://github.com/onnx/onnx}, 2019.

\bibitem{trt}
{NVIDIA}, \emph{{TensorRT}:
  \href{https://developer.nvidia.com}{https://developer.nvidia.com}},  2018.

\bibitem{cookiecutter_data_science}
{Driven Data}, ``Cookiecutter data science.''
  \url{https://drivendata.github.io/cookiecutter-data-science/}, 2022.

\bibitem{cookiecutter4fair}
{FAIR4HEP}, \emph{Cookiecutter4fair: v1.0.0},  2022.
\newblock 10.5281/zenodo.7306229.

\bibitem{hbb_interaction_network}
J.M.~Duarte, B.~Li, A.~Roy and R.~Zhu, \emph{{Hbb Interaction Network}:
  v0.1.1},  Nov., 2022.
\newblock 10.5281/zenodo.7305227.

\end{thebibliography}\endgroup
\bibliographystyle{JHEP}

\end{document}